\documentclass[twoside,slac_one]{revtex4}
\usepackage{graphicx}
\usepackage{fancyhdr}
\usepackage{amsmath} 
\usepackage{bm}
\usepackage{amsxtra}
\usepackage{amssymb}
\usepackage{amsthm}
\usepackage{latexsym}
\usepackage{lscape}

\begin{document}

\def\TeV {\ensuremath{\mathrm{TeV}}}
\def\GeV {\ensuremath{\mathrm{GeV}}}
\def\invnb   {\ensuremath{\mbox{\,nb}^{-1}}}
\def\invpb   {\ensuremath{\mbox{\,pb}^{-1}}}
\def\nb   {\ensuremath{\,\mbox{nb}}}
\def\ub   {\ensuremath{\,\mu\mbox{b}}}
\def\um   {\ensuremath{\,\mu\mbox{m}}}
\def\tev  {\ensuremath{\mbox{\,TeV}}}
\def\gev  {\ensuremath{\mbox{\,GeV}}}
\def\gevc  {\ensuremath{\mbox{\,GeV}/c}}
\def\gevcc  {\ensuremath{\mbox{\,GeV}/c^2}}
\def\mev  {\ensuremath{\mbox{\,MeV}}}
\def\mevc  {\ensuremath{\mbox{\,MeV}/c}}
\def\mevcc  {\ensuremath{\mbox{\,MeV}/c^2}}
\def\tmu   {\ensuremath{\tt trackerMuon}}
\def\tmus   {\ensuremath{\tt trackerMuons}}
\def\cc   {\ensuremath{c\bar{c}}}
\def\pp {\ensuremath{\mathrm{pp}}}
\def\Pp {\ensuremath{\mathrm{p}}}
\def\JPsi {\ensuremath{\mathrm{J/\psi}}}
\def\BBs {\ensuremath{\mathrm{B}_{\text{s}}^{\text{0}}}}
\def\BBd {\ensuremath{\mathrm{B}^{\text{0}}}}
\def\BBu {\ensuremath{\mathrm{B}^{\text{+}}}}
\def\BBBs {\ensuremath{\bar{\mathrm{B}}_{\text{s}}^{\text{0}}}}
\def\Pgm {\ensuremath{\mu}}
\def\Pgmp {\ensuremath{\mu^+}}
\def\Pgmm {\ensuremath{\mu^-}}
\def\PKp {\ensuremath{K^+}}
\def\PKm {\ensuremath{K^-}}
\def\psimumu {\ensuremath{\JPsi\to\Pgmp\Pgmm}}
\def\psik {\ensuremath{\BBu\to \JPsi \mathrm{K}^+}}
\def\psikst {\ensuremath{\BBd\to \JPsi \mathrm{K}^{*0}}}
\def\psiks {\ensuremath{\BBd\to \JPsi \mathrm{K}^{0}_{s}}}
\def\psiphi {\ensuremath{\BBs\to \JPsi\ \phi}}
\def\phikk {\ensuremath{\phi \to \PKp \PKm}}
\def\MB {\ensuremath{M_{\text{B}}}}
\def\mB {\ensuremath{m_{\text{b}}}}
\def\PB {\ensuremath{\mathrm{B}}}
\def\dsdpt {\ensuremath{d\sigma/dp_\mathrm{T}^{\PB}}}
\def\dsdeta {\ensuremath{d\sigma/d\eta^{\PB}}}
\def\dsdy {\ensuremath{d\sigma/d y^{\PB}}}
\def\ptb {\ensuremath{p_\mathrm{T}^{\PB}}}
\def\pt {\ensuremath{p_\mathrm{T}}}
\def\pl {\ensuremath{p_\mathrm{L}}}
\def\etab {\ensuremath{\eta^{\PB}}}
\def\yb {\ensuremath{\left|y^{\PB}\right|}}
\def\ybb {\ensuremath{y^{\PB}}}
\newcommand\T{\rule{0pt}{2.6ex}}
\newcommand\B{\rule[-1.2ex]{0pt}{0pt}}

\title{Study of the \boldmath$\psiphi$ decay in pp collisions at \boldmath$\sqrt{s}=7$ TeV with the CMS detector}

%

\author{G. Cerizza}
\affiliation{Department of Physics and Astronomy, University of Tennessee, Knoxville, TN, USA}

\begin{abstract}
B-hadrons are an ideal tool for advancing our current understanding of the flavor sector of the Standard
Model (SM). The study of B-meson production and decays is one of the key physics themes at the Large Hadron Collider (LHC) 
thanks to the large production rate and the fact that B-hadrons are relatively easy to trigger on and identify due
to their long lifetime and decays to muons. This talk presents the cross section
measurement for the exclusive final state \psiphi\ and an evaluation of the decay
branching fraction from the previously published exclusive-B production cross sections. 
Besides probing the heavy quark properties for the first time at the LHC energy, these measurements 
are also important tools for understanding and calibrating the detector, giving input for Monte 
Carlo tuning, and providing results for direct comparison with other experiments.
\end{abstract}

\maketitle

\thispagestyle{fancy}


\section{Heavy-flavor physics at the CMS experiment}

The measurements of differential cross sections for heavy-quark
production in high-energy hadronic interactions are critical input
for the underlying next-to-leading order (NLO) Quantum Chromodynamics
(QCD) calculations~\cite{NDE}.
While progress has been achieved in the understanding of heavy-quark
production at Tevatron energies~\cite{TeVI-CDF1,TeVI-CDF3,TeVI-CDF6,TeVI-D01,TeVI-D03,TeVI-D04,TeVII-CDF1,TEVII-CDF2,Cacciari04}, 
large theoretical uncertainties remain due to the dependence on the renormalization and factorization scales.
Measurements of b-hadron production at the higher energies provided by
the LHC represent an important new test of theoretical approaches that aim
to reduce the scale dependence of NLO QCD calculations~\cite{Cacciari98,Kniehl08}.
The Compact Muon Solenoid (CMS) experiment, that covers a rapidity
range complementary to the specialised b-physics experiment
LHCb~\cite{CMSLHCb}, recently measured the cross sections
for production of \BBu \cite{david} and \BBd \cite{keith}
in \pp\ collisions at $\sqrt{s}=7$ TeV. This talk presents the first measurement of the 
production of \BBs , with \BBs\ decaying into $\JPsi\  \phi$, that adds information to 
the understanding of b-quark production at this energy. Data and theoretical predictions are 
compared to NLO predictions of heavy-quark production.
The result, combined with the published \BBu\ and \BBd\ production cross section measurements, 
is then used for the evaluation of the \psiphi\ branching fraction. 
The cross section measurement for the \psiphi\ decay is only the first part 
of a longer term program that includes the measurement of the $B_s$ mesons properties 
(lifetime and lifetime difference) and CP-violating parameters (CP-even and CP-odd 
amplitude strenghts and weak phase).

\section{The CMS detector}

A detailed description of the CMS detector can be found elsewhere~\cite{JINST}.
The primary components used in this analysis are the silicon tracker and the muon
systems. The tracker operates in a $3.8$\,T axial magnetic field 
generated by a superconducting solenoid having an internal diameter of $6$~m.
The tracker consists of three cylindrical layers of silicon
pixel detectors complemented by two disks in the forward and backward
directions. The radial region between 20 and 116~cm is occupied
by several layers of silicon strip detectors in barrel and disk
configurations, ensuring at least nine hits in the pseudorapidity
range $|\eta| < 2.4$, where $\eta = -\ln{[\tan{(\theta/2)}]}$
and $\theta$ is the polar angle of the track measured from the positive $z$-axis of
a right-handed coordinate system, with the origin at the nominal interaction point,
the $x$-axis pointing to the centre of the LHC, the $y$-axis pointing up (perpendicular
to the LHC plane), and the $z$-axis along the counterclockwise-beam direction.
An impact parameter resolution of about 15\,$\mu$m and a \pt\ resolution of about 1.5\,\% are achieved
for charged particles with transverse momenta up to $100$~GeV/$c$.
Muons are identified in the range $|\eta|< 2.4$, with detection planes made of
drift tubes, cathode strip chambers, and resistive plate chambers,
embedded in the steel return yoke.
The first level of the CMS trigger system uses information from the crystal electromagnetic calorimeter, 
the brass/scintillator hadron calorimeter, and the muon detectors to select the most interesting events 
in less than 1 $\mu$s. The high level trigger (HLT) employs software algorithms and a farm of commercial 
processors to further decrease the event rate using information from all detector subsystems. 
The events used in the measurement reported in this paper were collected with a trigger requiring the presence 
of two muons at the HLT, with no explicit momentum threshold.

\section{Muon reconstruction}

In CMS muons are defined as tracks reconstructed in the silicon trackers and associated to a
compatible signal in the muon chambers. Two different muon types are available in CMS.
The first one, referred to as a \textit{Global Muon}, provides high-purity reconstruction for muons
with $\pt > 4$ GeV/c in the central pseudo-rapidity region $|\eta|<1.5$, and $\pt > 1$ GeV/c in the forward region.
\textit{Global Muons} are built as a combined fit of silicon and muon-chamber hits, belonging to
independent tracks found in the tracker and muon systems.
The second muon type, referred to as a \textit{Tracker Muon}, achieves a better reconstruction efficiency
at lower momenta. The requirements for a \textit{Tracker Muon} are relaxed compared to the \textit{Global Muons},
at the expense of a slightly larger background: tracks found in the tracker matched to only one
muon segment are accepted and not refitted. If two (or more) tracks are close to each other, it is
possible that the same muon segment or set of segments is associated to more than one track. In
this case the best track is selected based on the matching between the extrapolated track and the
segment in the muon detectors. 

\section{Strategy outline}

A sample of exclusive \psiphi\ decays, with \psimumu\ and \phikk , is reconstructed from the 
data collected in 2010 by the CMS experiment, corresponding to an integrated luminosity of
$39.6\pm 1.6\invpb$. The differential production cross sections, $\dsdpt$ and $\dsdy$, are
determined as functions of the transverse momentum \ptb\ and rapidity \yb\ of the
reconstructed \BBs\ candidate.
Here, the rapidity \ybb\ is defined as $\frac{1}{2}\ln{\frac{E+c\pl}{E-c\pl}}$, where $E$
is the particle's energy and $\pl$ is the particle's momentum along the counterclockwise
beam direction. The differential cross sections are calculated from the measured signal
yields ($n_{\rm sig}$), corrected for the overall efficiency ($\epsilon$), 
bin size ($\Delta x$, with $x =  \ptb ,\yb$), and integrated luminosity ($L$) as
\begin{equation}
\frac{d\sigma(\pp\to \psiphi)}{d x} = \frac{n_{\rm sig}}
{2\cdot \epsilon \cdot{\cal B}\cdot L\cdot\Delta x}\,\, ,
\label{eq:eq1}
\end{equation}
where ${\cal B}$ is the product of the branching fractions for the decays of the $\JPsi$ and $\phi$ mesons.
In each bin the signal yield is extracted with an unbinned maximum likelihood fit to the
$\JPsi\ \phi$ invariant mass and proper decay length $ct$ of the \BBs\ candidates.
The factor of two in Eq.~\ref{eq:eq1} is required since we report the result as a cross 
section for \BBs\ production alone, while both \BBs\ and \BBBs\ are included in $n_{\rm sig}$.
The size of the bins is chosen such that the statistical uncertainty on $n_{\rm sig}$ is comparable in each of them.


\subsection{Reconstruction and \boldmath$\BBs$ selection}

Reconstruction of \psiphi\ candidates begins by identifying \psimumu\ decays.
The muon candidates must have one or more reconstructed segments in the muon system that match the extrapolated position of a track
reconstructed in the tracker. Furthermore, the muons are required to
lie within a kinematic acceptance region defined
as: $\pt^{\mu}> 3.3$~GeV/$c$ for $|\eta^{\mu}| < 1.3$;
total momentum $p^{\mu}> 2.9$~GeV/$c$ for $1.3 < |\eta^{\mu}| < 2.2$;
and $\pt^{\mu}> 0.8$~GeV/$c$ for $2.2 < |\eta^{\mu}| < 2.4$.
Two oppositely charged muon candidates are paired and are required to originate from a
common vertex using a Kalman vertex fit~\cite{kalman}.
The muon pair is required to have a transverse momentum $\pt > 0.5$~GeV/$c$
and an invariant mass within 150~MeV/$c^2$ of the world average $\JPsi$ mass value~\cite{PDG2010},
which corresponds to more than three times the measured dimuon invariant mass resolution~\cite{BPH-10-002}.

Candidate $\phi$ mesons are reconstructed from pairs of oppositely
charged tracks with $\pt>0.7$~GeV/$c$ that are selected from a
sample with the muon candidate tracks removed.
The tracks are required to have at least five hits in the silicon tracker detectors,
and a track $\chi^2$ per degree of freedom less than five.
Each track is assumed to be a kaon and the invariant mass
of a track pair has to be within $10$~MeV/$c^2$
of the world average $\phi$-meson mass~\cite{PDG2010}.

The \BBs\ candidates are formed by combining a $\JPsi$ candidate with a $\phi$ candidate.
The two muons and the two kaons are subjected to a combined vertex and kinematic fit~\cite{kinfit}, where in addition
the dimuon invariant mass is constrained to the nominal $\JPsi$ mass.
The selected candidates must have a resulting $\chi^2$ vertex probability
greater than $2\%$, an invariant mass between $5.20$ and $5.65$~GeV/$c^2$,
and be in the kinematic range $8< \ptb <50$~GeV/$c$ and $\yb<2.4$.
For events with more than one candidate, the one with the highest
vertex-fit probability is selected. This results in the correct
choice 97\% of all cases, as determined from simulated signal events.

The proper decay length of each selected \BBs\ candidate
is calculated using the formula $ct = c(\MB/ \ptb)L_{xy}$, where the transverse decay
length $L_{xy}$ is the length of the vector $\vec{s}$ pointing from the primary
vertex~\cite{TRK-10-005} to the secondary vertex projected onto the
\BBs\ transverse momentum:
$L_{xy} = (\vec{s}\cdot\vec{p}_T^{\PB})/{\ptb}$, with $\MB$ the reconstructed mass
of the \BBs\ candidate.
Candidate \BBs\ mesons are accepted within the range $-0.05 < ct < 0.35$~cm.
A total of $6,200$ events pass all the selection criteria.

The efficiency of the \BBs\ reconstruction is computed with a combination of techniques using the data and
large samples of simulated signal events generated using PYTHIA\ $6.422$~\cite{PYTHIA}.
The decays of unstable particles are described by the EVTGEN~\cite{EvtGen} simulation.
Long-lived particles are then propagated through a detailed description of the CMS detector based on the GEANT4~\cite{GEANT4}
package. The trigger and muon-reconstruction efficiencies are obtained from a large sample of inclusive
$\psimumu$ decays. The total efficiency of this selection, defined as the fraction of \psiphi\
decays produced with $8< \ptb <50$~GeV/$c$ and $\yb<2.4$ that pass all criteria, ranges from
1.3\% for $\ptb\approx\,8$~GeV/$c$ to $19.6\%$ for $\ptb>23$~GeV/$c$.

\subsection{Muon efficiency}

The ``Tag and Probe''~\cite{BPH-10-002} technique is a data driven method to measure the single 
muon tracking, identification and trigger efficiencies. It makes use of a well-known dimuon resonance (such as \JPsi\ mesons) to supply 
tags and probes. The choice of such resonance is due to the CMS experiment ability to measure muon momenta
with high precision and reconstruct and identify muons with high efficiencies. 
Events are selected with strict selection requirements on one muon ({\it tag}), and with a more relaxed selection 
on the other track ({\it probe}), such that the selection applied to the probe track does not bias the efficiency that one 
wants to measure. The {\it probe} tracks are separated into two categories depending on whether they pass or fail 
the more restrictive selection. 
If $p_{tag}$ and $p_{probe}$ are the four-momenta of the {\it tag} muon and {\it probe} track respectively then the 
invariant mass m of the particle is given by:
\begin{equation}
m = \sqrt{p_{tag}+p_{probe}}
\end{equation}
The technical chain runs as follows: all events of the \JPsi\ samples are passed through. 
An event is kept if a tag muon $+$ probe track combination is found while satisfying the predefined criteria. 
In this case the invariant mass of the combination enters a muon-track mass histogram. The criteria for the choice of both, {\it tag} and {\it probe},
should be defined such that background is reduced and the \JPsi\ mass peak is clearly visible.
If the track is subsequently identified as a global/tracker muon the invariant mass of the tag and 
probe combination enters another muon-muon mass histogram. The latter should contain almost exclusively $\JPsi\to\mu^+\mu^-$ events.
A simultaneous unbinned Maximum Likelihood (ML) fit to both mass distribution extracts the efficiency.
The dimuon efficiencies are calculated as the product of the single-muon efficiencies obtained with this method.
Corrections to account for correlations between the two muons (1--3\%) are obtained from simulation studies.
The correction factors are determined in bins of single muon $\pt^{\mu}$ and $\eta^{\mu}$, and are applied
independently to each muon from a \psiphi\ decay in the simulation to determine the total corrected efficiency.


\section{Fit technique}

The two main background sources are prompt and non-prompt $\JPsi$ production.
The latter background is mainly composed of \BBu\ and \BBd\ mesons that decay to a
\JPsi\ and a higher-mass K-meson state (such as the $\mathrm{K}^{\mathrm{+}}_{\mathrm{1}}$). Such events
tend to have lower reconstructed \MB\ mass.
Inspection of the reconstructed $\JPsi\phi$ invariant mass for a large variety of potential B background channels confirms that
there is no single dominant component and that the channel
$\BBd\to \JPsi\ \mathrm{K}^{\mathrm{*0}}$ (with $\mathrm{K}^{\mathrm{*0}}\to \PKp\pi^-$),
which \textit{a priori} is kinematically similar to the signal decay and more
abundantly produced, is strongly suppressed by the restriction on the $\PKp\PKm$ invariant mass.
A study of the sidebands of the dimuon invariant mass distribution confirms that
the contamination from events without a $\JPsi$ decay to two muons is negligible after all selection
criteria have been applied.

The signal yields in each \ptb\ and \yb\ bin are obtained using an unbinned extended maximum-likelihood fit to $\MB$ and $ct$.
The likelihood for event $j$ is obtained by summing the product of the yield $n_i$ and
the probability density functions (PDF) ${\cal P}_i$ and ${\cal Q}_i$ for each of the signal and background hypotheses
$i$.  Three individual components are considered: signal,
non-prompt b $\to\JPsi$ X, and prompt $\JPsi$.  The extended likelihood function
is then the product of likelihoods for each event $j$:
\begin{equation}
{\cal L} = \exp\left ( -\sum_{i=1}^{3} n_i \right ) \prod_j \left [
\sum_{i=1}^{3} n_i{\cal P}_i(\MB;\, \vec{\alpha}_i){\cal Q}_i(ct;\, \vec{\beta}_i) \right ].
\end{equation}
The PDFs ${\cal P}_i$ and ${\cal Q}_i$ are parameterized separately for each fit component
with shape parameters $\vec{\alpha}_i$ for $\MB$ and $\vec{\beta}_i$ for $ct$.
The yields $n_i$ are then determined by minimizing the quantity $-\ln{\cal L}$ with
respect to the signal yields and a subset of the PDF parameters~\cite{roofit}.
The PDFs are constructed from basic analytical functions that satisfactorily describe
the variable distributions from simulated events. Shape parameters are
obtained from data when possible. The $\MB$ PDF is the sum of two
Gaussian functions for the signal, a second-order polynomial for the non-prompt
$\JPsi$ that allows for possible curvature in the shape, and a first-order
polynomial for prompt $\JPsi$. The resolution on $\MB$ is approximately $20$~MeV/$c^2$ near the \BBs\ mass.

For the signal, the $ct$ PDF is a single exponential parameterized in terms of
a proper decay length $c\tau$. It is convolved with a resolution
function that is a combination of two Gaussian functions to account for
a dominant core and small outlier distribution; the core fraction is
varied in the fit and found to be consistently larger than 95\%.
The $ct$ distribution for the non-prompt $\JPsi$ background is described
by a sum of two exponentials, with effective lifetimes that are allowed
to be different. The ``long-lifetime exponential'' corresponds to decays of b-hadrons to
a \JPsi\ plus some charged particles that survive the $\phi$ selection, while the
``short-lifetime exponential'' accounts for events where the muons from the
\JPsi\ decay are wrongly combined with hadron tracks originating from the
\pp\ collision point. The exponential functions are convolved with a resolution function
with the same parameters as the signal.
For the prompt $\JPsi$ component the pure resolution function is used.
The core resolution in $ct$ is measured in data to be $45\ \mu$m.
All background shapes are obtained directly from data, while the signal shape in $\MB$ is taken from
a fit to reconstructed signal events from the simulation.
The effective lifetime and resolution function parameters for prompt and non-prompt
backgrounds are extracted from data in regions of $\MB$ that are separated
by more than four times the width of the observed \BBs\ signal from the mean
\BBs\ peak position ($\MB$ sidebands): $5.20 < \MB\ < 5.29$~GeV/$c^2$ and
$5.45 < \MB\ < 5.65$~GeV/$c^2$. A comparison of the PDF shapes for the different
sideband regions in simulated events confirms that they are extrapolated well into the signal region.
With the lifetimes for signal and non-prompt background fixed from this first step, the
resolution function parameters are then determined separately in each
\ptb\ and \yb\ bin, from the $\MB$ sidebands.
The signal and background yields in each \ptb\ and \yb\  bin are determined
in a final iteration, using the full \MB\ range, with all parameters floating
except the background lifetimes and the lifetime resolution functions, which
are fixed to the results of the fit to the \MB\ sidebands.
It has been verified that leaving all parameters floating
changes the signal yield by an amount smaller than the systematic
uncertainty assigned to the fit procedure.


\subsection{Fit results}

\begin{figure}[h]
\centering
\includegraphics[width=135mm]{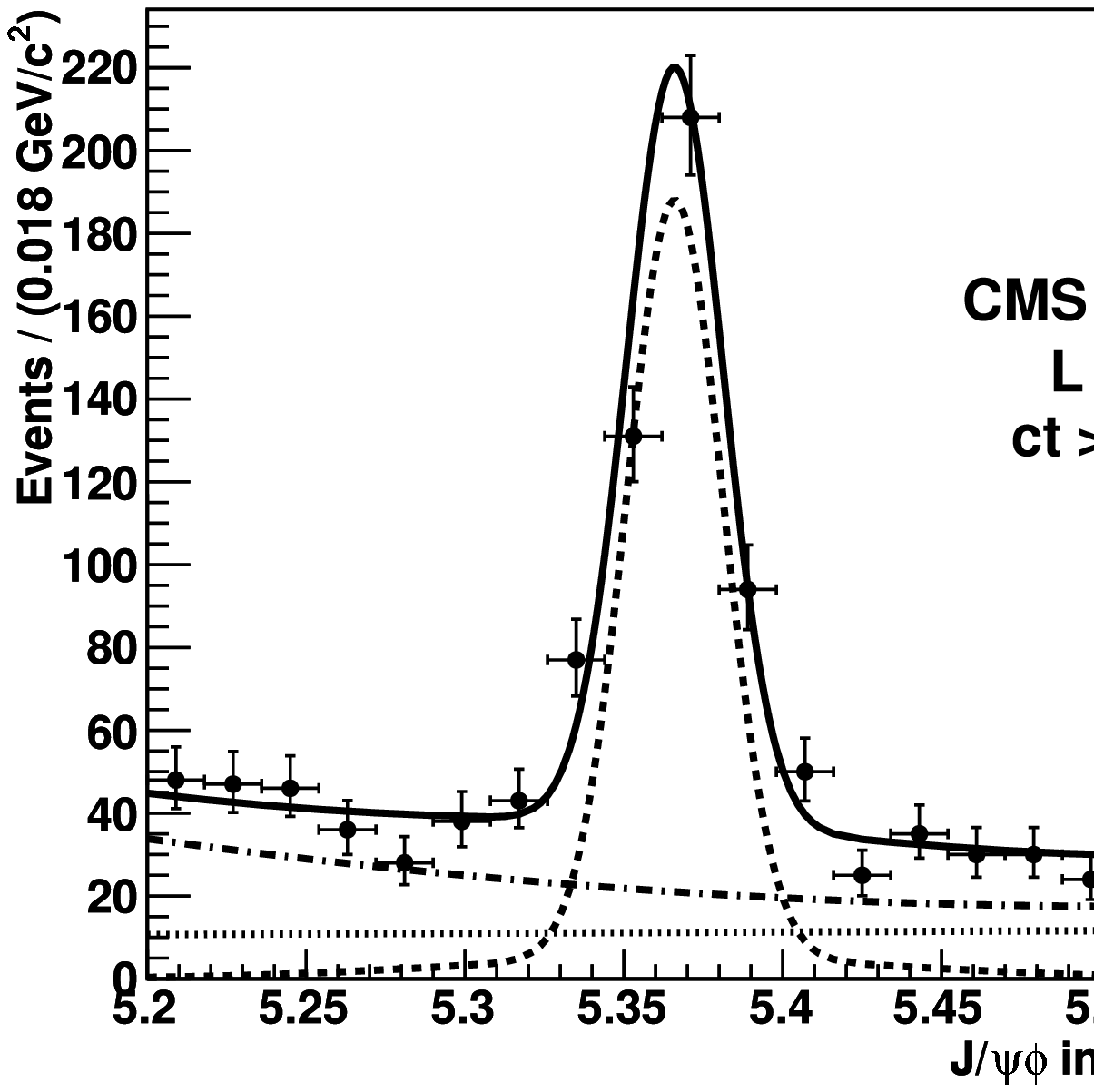}
\includegraphics[width=135mm]{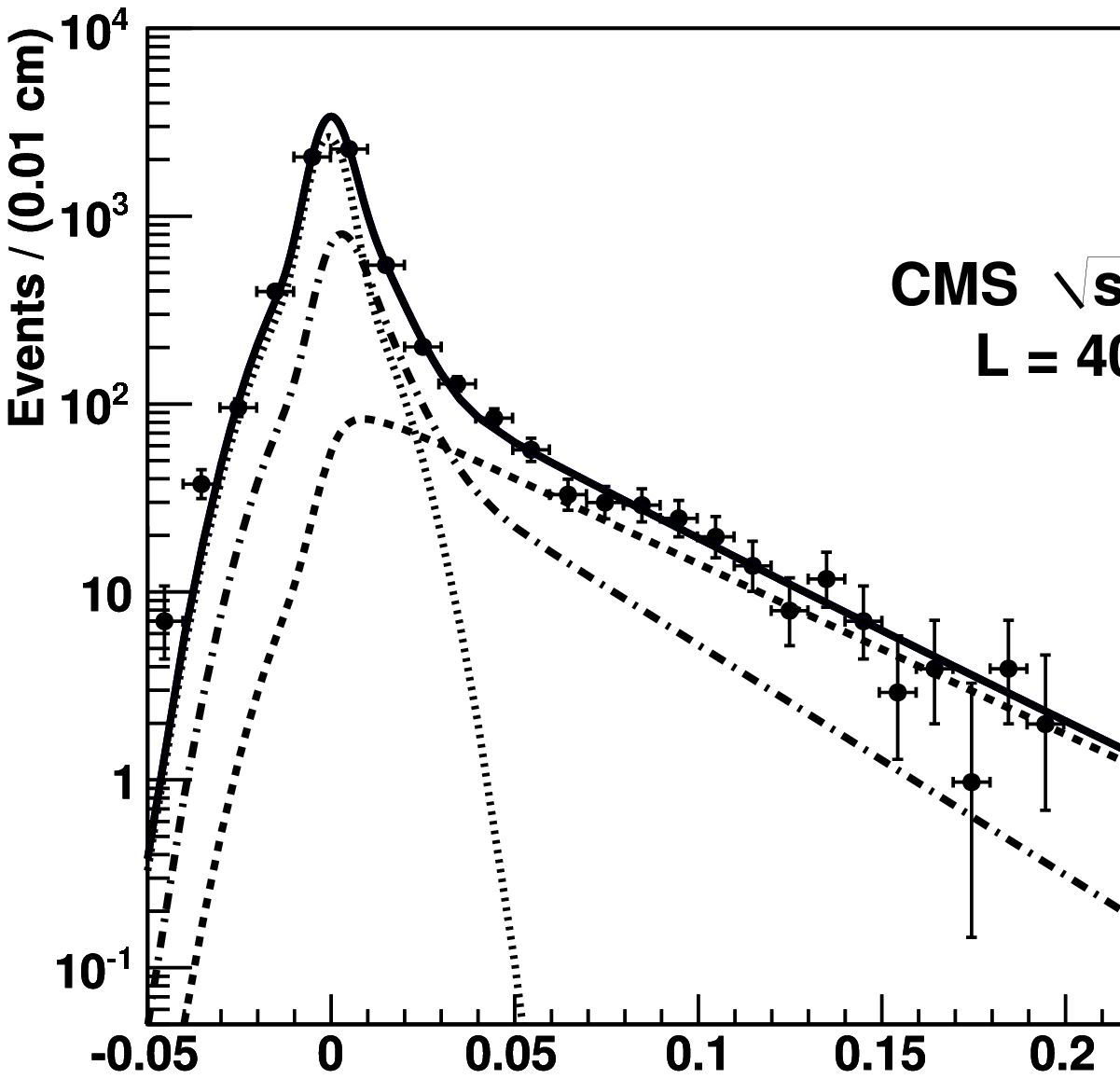}
\caption{Projections of the fit results in $\MB$ (a) and $ct$ (b) for
  $8< \ptb <50$~GeV/$c$ and $\yb < 2.4$. The curves in each plot are: the sum of all
  contributions (solid line); signal (dashed); prompt $\JPsi$ (dotted);
  and non-prompt $\JPsi$ (dot-dashed).
  For better visibility of the individual contributions, plot (a) includes the requirement $ct>0.01\ {\rm cm}$.} 
\label{fit}
\end{figure}

Figure~\ref{fit} shows the fit projections for $\MB$ and $ct$ from the
inclusive sample with $8< \ptb <50$~GeV/$c$ and $\yb < 2.4$.  When plotting $\MB$,
the selection $ct>0.01$ cm is applied for better visibility of the individual
contributions. The number of signal events in the entire data sample is $549\pm 32$, where the uncertainty is
statistical only. The obtained proper decay length of the signal, $c\tau = 478 \pm 26$ $\mu$m,
is within $1.4$ standard deviations of the world average value~\cite{PDG2010}, even though
this analysis was not optimized for lifetime measurements.

\section{Systematics}

The cross section measurement is affected by several sources of systematic uncertainty arising
from uncertainties on the fit, efficiencies, branching fractions,
and integrated luminosity. In every bin the total uncertainty is about $11\%$.
Uncertainties on the muon efficiencies from the trigger, identification, and tracking
are determined directly from data ($3-5\%$). The uncertainty of the method
employed to measure the efficiency in the data has been
estimated from a large sample of full-detector simulated events ($1-3\%$).
The tracking efficiency for the charged kaons is
consistent with simulation. A conservative uncertainty of at most $9\%$
in each bin has been assigned for the hadronic track reconstruction (adding linearly
the uncertainties on the two kaon tracks~\cite{TRK-10-002}), which includes the
uncertainty due to misalignment of the silicon detectors.
The uncertainty of the fit procedure arising from potential
biases and imperfect knowledge of the PDF parameters is estimated by varying the parameters
by one standard deviation ($2-4\%$).
The contribution related to the \BBs\ momentum spectrum ($1-3\%$) is evaluated
by reweighting the shape of the \ptb\ distribution generated with PYTHIA to match
the spectrum predicted by MC@NLO~\cite{MCNLO}. An uncertainty of $1\%$ is assigned to the
variation of the selection criteria applied to the vertex fit probability, the
transverse momentum of the kaons, the \BBs\ transverse momentum, and the $\PKp\PKm$ invariant mass.
An uncertainty is added to account for the limited number of simulated events
(at most $3\%$ in the highest \ptb\ bin).
The total uncorrelated systematic uncertainty in each bin is
the sum in quadrature of the individual uncertainties (see Table~\ref{tablesyst}).
In addition, there are common uncertainties of $4\%$ from the integrated luminosity measurement~\cite{EWK-10-004} and
$1.4\%$ from the $\JPsi$ and $\phi$ branching fractions~\cite{PDG2010}.
As the reported result is a measurement of the \BBs\ cross section times the
\psiphi\ branching fraction, the $30\%$ uncertainty in the
\psiphi\ branching fraction~\cite{PDG2010} is not included in the result.

\begin{table}[h]
\begin{center}
\caption{Summary table of the relative systematic uncertainties in the measurement of the \psiphi\ 
production cross section.}
\begin{tabular}{|l|c|}
\hline \textbf{Source} & \textbf{Uncertainty ($\%$)} \\ \hline 
Muon Reconstruction Efficiency & $3-5$ \\
Hadron Tracking Efficiency & $7.8$ \\
Reconstruction Efficiency & $2-3$ \\
Misalignment & $2-4$ \\
\ptb -\yb\ Spectrum & $1-3$ \\
Probability Density Function & $2-4$ \\ \hline
Uncorrelated Systematic Errors & $10-11$ \\ \hline
Branching Fractions & $1.4$ \\
Luminosity & $4$ \\ \hline
Correlated Systematic Errors & $4.2$ \\ \hline
Total Systematic Error & $11-12$ \\ \hline
\end{tabular}
\label{tablesyst}
\end{center}
\end{table}


\section{Differential cross section measurement}

\begin{figure}[h]
\centering
\includegraphics[width=135mm]{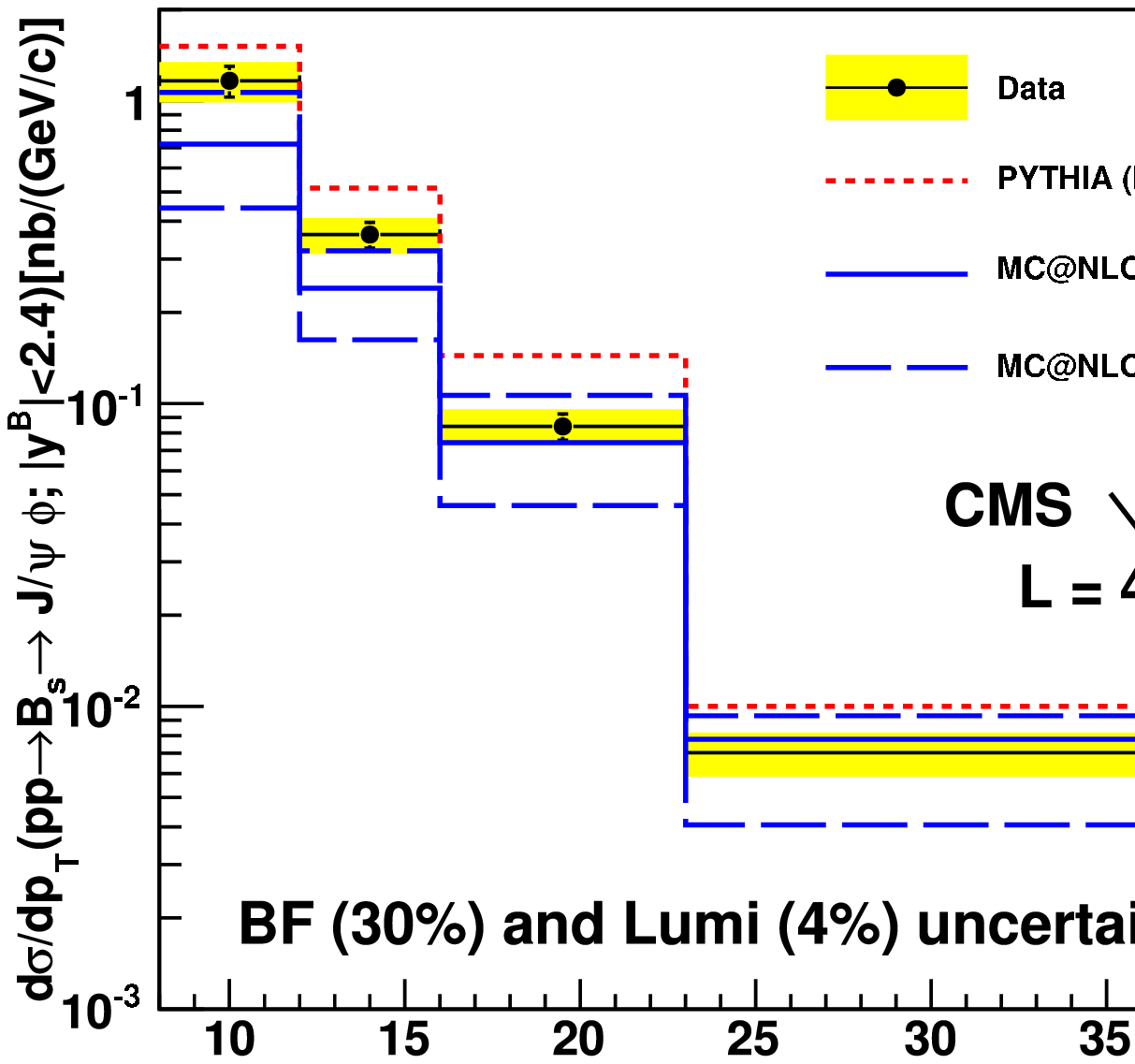}
\includegraphics[width=135mm]{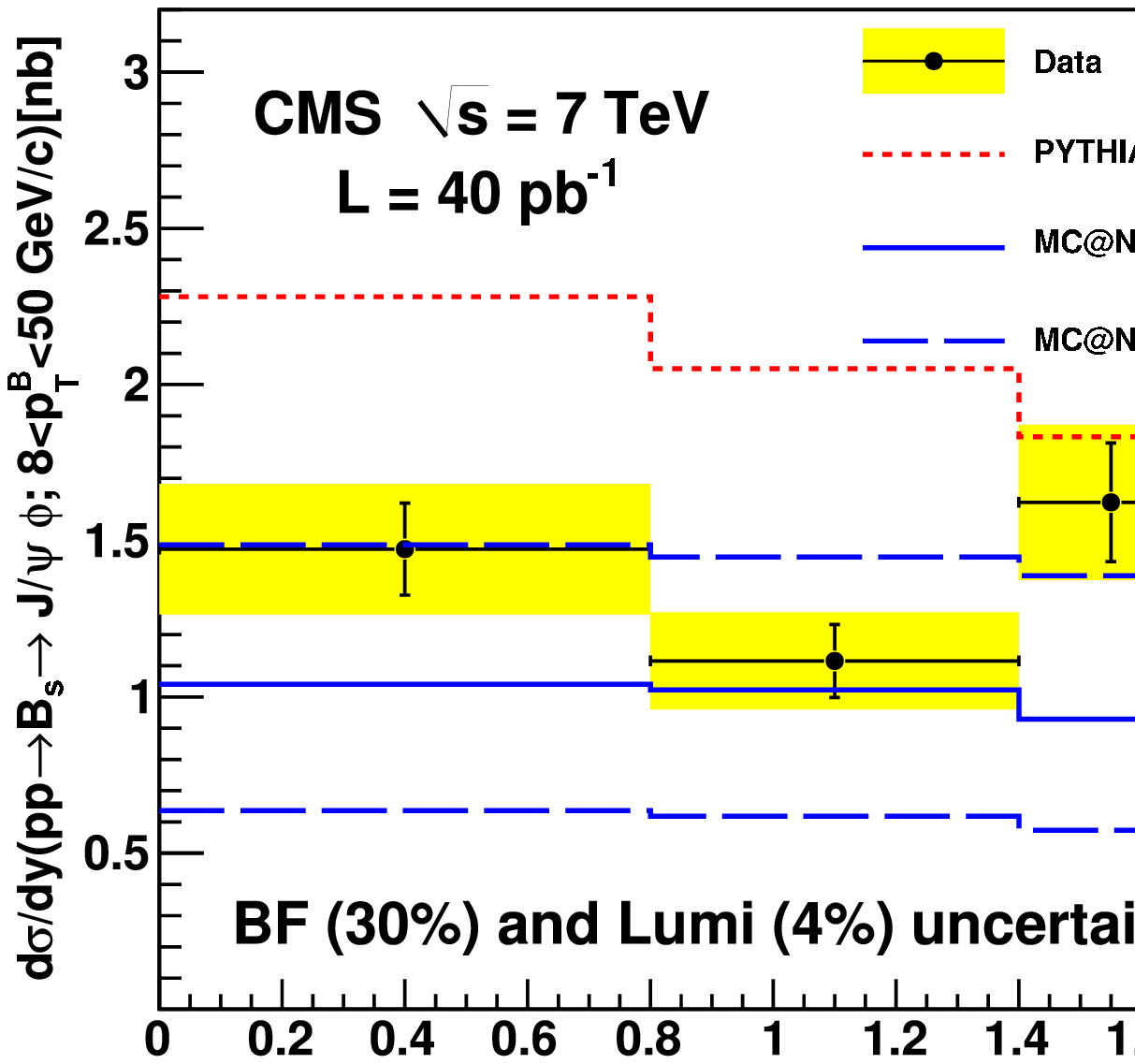}
\caption{Measured differential cross sections $\dsdpt$ (a) and $\dsdy$ (b) compared with
theoretical predictions.  The (yellow) band represents the sum in
quadrature of statistical and systematic uncertainties.
The dotted (red) line is the PYTHIA prediction; the solid and
dashed (blue) lines are the MC@NLO prediction and its uncertainty, respectively.
The common uncertainties of $4\%$ on the data points, due to the integrated luminosity,
and of $30\%$ on the theory curves, due to the \psiphi\ branching fraction, are not shown.}
\label{fig:Xsec}
\end{figure}

The differential cross sections times branching fraction as functions of \ptb\ and \yb\ are 
plotted in Fig.~\ref{fig:Xsec}, together with predictions from MC@NLO and PYTHIA .
The predictions of MC@NLO use the renormalization and factorization scales $\mu=\sqrt{\mB^2c^4+\pt^2c^2}$, where \pt\ is
the transverse momentum of the b quark, a b-quark mass of $\mB = 4.75$~GeV/$c^2$, and the CTEQ6M parton distribution
functions~\cite{CTEQ}. The uncertainty in the MC@NLO cross section is obtained simultaneously varying the renormalization and
factorization scales by factors of two, varying \mB\ by $\pm 0.25$~GeV/$c^2$, and using the CTEQ6.6 parton distribution
function set. The prediction of PYTHIA uses the CTEQ6L1 parton distribution
functions~\cite{CTEQ}, a b-quark mass of $4.8$~GeV/$c^2$, and the Z2 tune~\cite{Z2} to
simulate the underlying event.
The total integrated \BBs\ cross section times \psiphi\ branching fraction for the
range $8 < \ptb < 50$~GeV/$c$ and $\yb < 2.4$ is
measured to be ($6.9\pm 0.6\pm 0.6$)~nb, where the first uncertainty is statistical and the second is systematic.
The statistical and systematic uncertainties are derived from the bin-by-bin uncertainties and propagated through the sum.
The measured total cross section lies between the theoretical predictions of MC@NLO ($4.6^{+1.9}_{-1.7}\pm 1.4$~nb)
and PYTHIA ($9.4\pm2.8$~nb), where
the last uncertainty is from the \psiphi\ branching fraction~\cite{PDG2010}.
Also the previous CMS cross-section measurements
of $\BBu$~\cite{david} and $\BBd$~\cite{keith} production in \pp\ collisions at $\sqrt{s}=7$ TeV,
gave values between the two theory predictions, indicating internal consistency amongst the three
different B-meson results.

\section{Evaluation of the branching fraction for \boldmath$\psiphi$}

The branching fraction BF(\psiphi) can be  calculated independently with respect to either the \psik\ 
or the \psiks\ decay channels. The observed cross section for the decay mode \psiphi\ can be written as:
\begin{equation}
\sigma(pp\to \psiphi) = \sigma(pp\to \bar{b})
\cdot f_s \cdot BF(\psiphi)\cdot f_{kin}^{\BBs}
\label{eq:y}
\end{equation}
and similarly, for the \BBu\ and \BBd\ mode:
\begin{eqnarray}
\sigma(pp\to \BBu X) & = & \sigma(pp\to \bar{b})
\cdot f_u \cdot f_{kin}^{\BBu} \\
\sigma(pp\to \BBd X) & = & \sigma(pp\to \bar{b})
\cdot f_d \cdot f_{kin}^{\BBd}
\end{eqnarray}
Here, the $f_u$, $f_d$, and $f_s$ are the probabilities that the b anti-quark will hadronize and form a \BBu, \BBd, and \BBs\ meson, 
respectively. The fractions $f_{kin}^\mathrm{B}$ correct for the limited range in rapidity and transverse 
momentum in the different analyses. The extrapolation to the full kinematic range is theory dependent.
The NLO theory predictions~\cite{MCNLO} for the expected differential cross section~\cite{david,keith,giordano} 
values are in good agreement with the measured ones in each of the three decay channels. 
Therefore, it is possible to identify the best central model (CTEQ6M, $Q_R=Q_F=1$, and $\mB = 4.75$~GeV/$c^2$) 
to predict the full kinematic range in \ptb\ and \yb. The spectra are obtained from a number of NLO generated events large enough
to keep the relative statistical error on the predicted ratio at 0.1\%. 
The values for the branching fractions are derived from the measured cross sections:
\begin{eqnarray}
\mathrm{BF}(\psiphi) & = & 
\frac{\sigma(pp\to \psiphi)}{\sigma(pp\to \mathrm{B^{+,0}} X)}
\cdot \frac{f_{u,d}}{f_s}\cdot \frac{f_{kin}^{\mathrm{B^{+,0}}}}{f_{kin}^{\BBs}}
\label{eq:cross}
\end{eqnarray}
Both values depend on the ratio of the fraction of kinematic ranges that vary slowly 
over a wide range of model parameters. 
The values for the fragmentation fractions are obtained from HFAG, printed 
in PDG~\cite{PDG2010} under b-hadron admixtures.
Both results, from Tevatron and from the combination
of LEP and Tevatron measurements, are considered and reported in Table~\ref{tab:frag}. 
The sources of uncertainties are listed in Table~\ref{tab:error}.

\begin{table*}[t]
\begin{center}
\caption{Summary table for the fragmentation fractions used in the evaluation of
the \psiphi\ branching fraction.}
\begin{tabular}{|l|c|}
\hline
\multicolumn{2}{|c|}{\textbf{LEP+Tevatron}} \\
\hline 
$\Gamma(\bar{b}\rightarrow \BBs)$ & $(11.0 \pm 1.2) \%$ \\
$\Gamma(\bar{b}\rightarrow \BBu)$ & $(40.3 \pm 1.1) \%$ \\
$\Gamma(\bar{b}\rightarrow \BBd)$ & $(40.3 \pm 1.1) \%$\\ \hline
$\frac{f_s}{f_{u,d}}=\frac{\Gamma(\BBs)}{\Gamma(\mathrm{B^{+,0}})}$ & $(27.2\pm 3.1) \%$   \\
\hline 
\multicolumn{2}{|c|}{\textbf{Tevatron}} \\
\hline 
$\Gamma(\bar{b}\rightarrow \BBs)$ & $(11.1 \pm 1.4) \%$ \\
$\Gamma(\bar{b}\rightarrow \BBu)$ & $(33.9 \pm 3.1) \%$ \\
$\Gamma(\bar{b}\rightarrow \BBd)$ & $(33.9 \pm 3.1) \%$\\ \hline
$\frac{f_s}{f_{u,d}}=\frac{\Gamma(\BBs)}{\Gamma(\mathrm{B^{+,0}})}$ & $(32.7\pm 5.1) \%$   \\
\hline 
\end{tabular}
\label{tab:frag}
\end{center}
\end{table*}

\begin{table}[h]
\begin{center}
\caption{Summary table of the systematic uncertainties for the evaluation of the \psiphi\ 
branching fraction.}
\begin{tabular}{|l|c|c|}
\hline \textbf{Source} & \textbf{$\mathrm{BF}^{\psiphi}_{\psik}$} & \textbf{$\mathrm{BF}^{\psiphi}_{\psiks}$} \\ \hline 
& \multicolumn{2}{|c|}{Experimental Uncertainties} \\ \hline
Cross Section & $15.8$ & $16.5$ \\
NLO Spectrum & $4.6$ & $4.3$ \\  \hline
& \multicolumn{2}{|c|}{PDG Uncertainties} \\ \hline
Branching Fractions & $3.5$ & $3.8$ \\
Fragmentation Fractions & $11.2$ & $11.2$ \\ \hline
\end{tabular}
\label{tab:error}
\end{center}
\end{table}

The experimental error is the sum in quadrature
of the statistical and systematic uncertainties calculated as described 
in the cross section measurement papers~\cite{david,keith,giordano}. 
It contains the uncertainty of the yield as extracted from maximum likelihood fits, 
the uncertainty on the reconstruction and hadron-tracking efficiencies,  
misalignment, and the variation with different models to correct for 
the limited kinematic ranges. 
From that, we distinguish the uncertainties of the branching fractions 
and the fragmentation fractions in PDG~\cite{PDG2010}. 
Individual uncertainties are added in quadrature. 
The cross section measurements in the different exclusive B decay modes, omitting
the uncertainty due to the luminosity measurement, are:
\begin{itemize}
\item $\sigma(pp\to \psiphi) = (6.9 \pm 0.6 \pm 0.5)\times 10^{-3} \mu$b 
\item $\sigma(pp\to \BBu X) = (28.3 \pm 2.4 \pm 2.0) \mu$b
\item $\sigma(pp\to \BBd X) = (33.2 \pm 2.5 \pm 3.1) \mu$b
\end{itemize}
The branching fraction BF(\psiphi) is calculated according to
Eq.~\ref{eq:cross} with respect to the \BBu\ meson and to the \BBd\ meson productions.
The two measurements of the branching fraction BF(\psiphi)
overlap within one standard deviation. Their error weighted average is
\begin{equation}
\mathrm{BF}(\psiphi) = (1.8 \pm 0.2 \pm 0.2) \times 10^{-3}
\end{equation}
where the errors are the combined experimental and PDG uncertainties, respectively.
The result is listed in the same way as PDG does for
the present world average. The result agrees within one standard deviation with the
PDG value~\cite{PDG2010} of $\mathrm{BF}(\psiphi) = (1.4 \pm 0.4 \pm 0.2) \times 10^{-3}$
assuming the same uncertainty on the b-quark fragmentation fraction.
Using the fragmentation fractions as extracted from the Tevatron, only, 
the error weighted average for the branching fraction is
\begin{equation}
\mathrm{BF}(\psiphi) = (1.5 \pm 0.2 \pm 0.2) \times 10^{-3}
\end{equation}
to be compared with the PDG value $\mathrm{BF}(\psiphi) = (1.2 \pm 0.3 \pm 0.2) \times 10^{-3}$
that has been extracted in the same way. 


\section{Summary}

In summary, the first measurements of the \BBs\ differential cross sections $\dsdpt$ and $\dsdy$
in pp collisions at $\sqrt{s}=7$ TeV and in the decay channel \psiphi\ have
been presented.  The measurement has been performed in four bins in the kinematic range $\yb < 2.4$ and
$8<\ptb\ <50$~GeV/$c$. This study complements previous results in moving towards
a comprehensive description of b-hadron production at $\sqrt{s}=7$ TeV.
An estimation of the \psiphi\ branching fraction has been calculated from the published 
CMS measurements of inclusive \BBu\ and \BBd\ production cross section.
We calculate the branching fraction for the \psiphi\ decay, assuming the fragmentation fractions
extracted from measurements at the Tevatron, to $(1.5 \pm 0.2 \pm 0.2) \times 10^{-3}$ that agrees
within one standard deviation with the value published from the Tevatron.

\clearpage
\bigskip 
\bibliography{bibliography}

\end{document}